\documentclass{article}
\usepackage{glas}
\usepackage{epsfig}
\usepackage{subfigure}
\newcommand{\als}{$\alpha_s$}
\newcommand{\ee}{\mbox{$e^+e^-$}}
\newcommand{\logxp}     {\mbox{$\log(1/x_{p})$}}
\newcommand{\logxpmax}  {\mbox{$\log(1/x_{p})_{max}$}}
\begin{document}
\begin{titlepage}{GLAS-PPE/98--02}{May 1998}
\title{QCD at HERA}
\author{N.\ H.\ Brook}
\collaboration{on behalf of the ZEUS \& H1 collaborations}
\begin{abstract}
A review of HERA measurements of structure functions, fragmentation
functions and forward jet production is presented.
\end{abstract}
\vfill
\conference{invited talk given at the \\
 2nd Latin American Symposium \\
on High Energy Physics, \\
San Juan, Puerto Rico. \\Apr 8-11, 1998}
\end{titlepage}

\section{DIS kinematics}
The event kinematics of deep inelastic scattering, DIS,
are determined by the negative square of the four-momentum transfer at
the lepton vertex,
$Q^2\equiv-q^2$, and the Bjorken scaling variable, $x=Q^2/2P \cdot q$,
where $P$ is the four-momentum of the proton.
In the quark parton model (QPM),
the interacting quark from the proton carries the four-momentum $xP$.
The variable $y$, the fractional energy transfer to the proton in its
rest
frame, is related to $x$ and $Q^2$ by $y\simeq Q^2/xs$, where $\sqrt s$
is
the positron-proton centre of mass energy.
Because the H1 and ZEUS detectors are almost hermetic the kinematic
variables $x$ and $Q^2$ can be reconstructed in a variety of
ways using combinations of positron and hadronic system energies and
angles~\cite{DA}.

\begin{figure}[htb]
\centerline{\epsfig{file=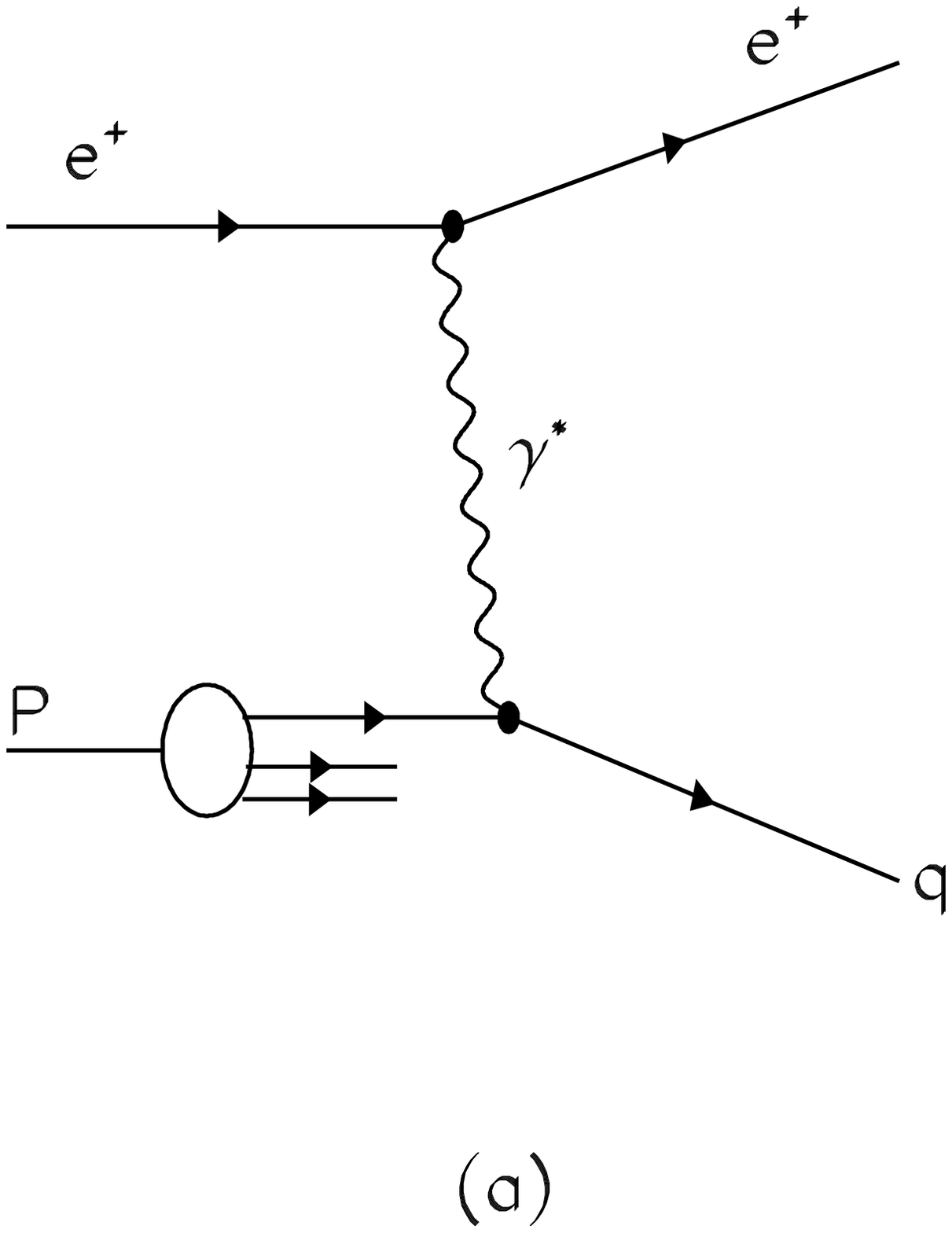,width=0.3\textwidth}
\epsfig{file=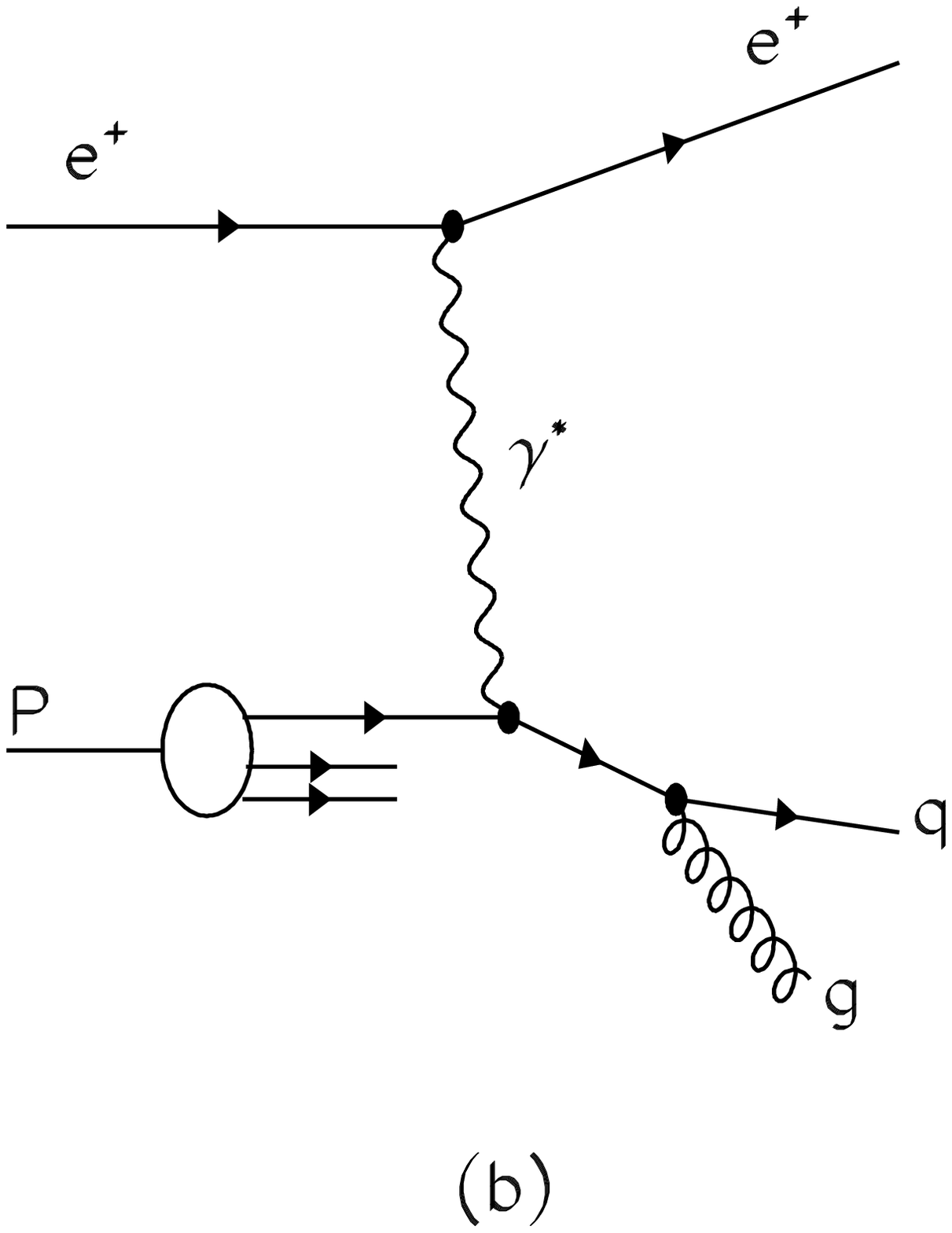,width=0.3\textwidth}
\epsfig{file=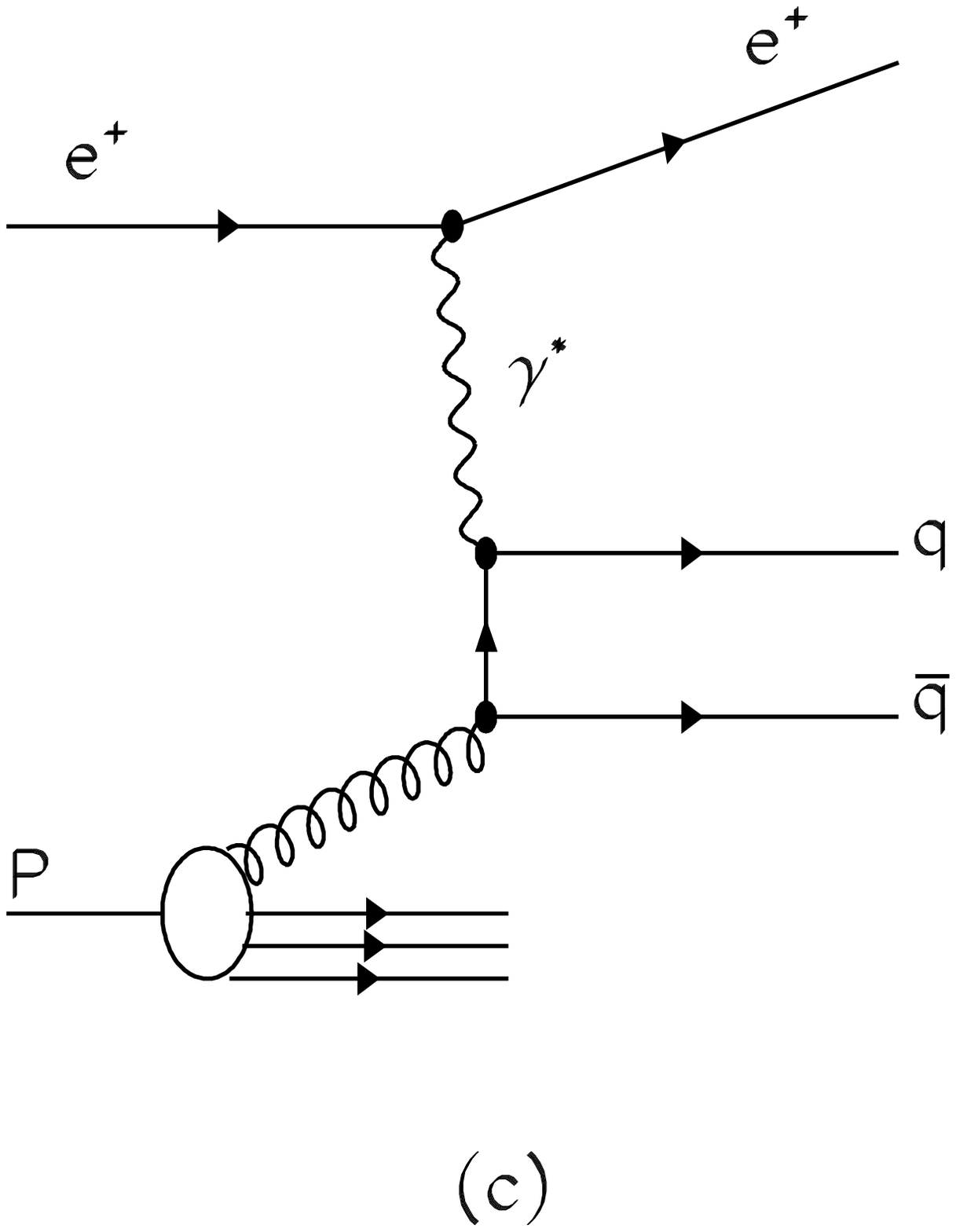,width=0.3\textwidth}}
\caption{(a) QPM (b) QCDC  and (c) BGF diagrams}
\label{fig:feyn_alphas}
\end{figure}

Neutral current (NC) DIS occurs when an uncharged boson ($\gamma, Z^0$)
is exchanged between the lepton and proton.
In QPM there is a
1+1 parton configuration, fig.~\ref{fig:feyn_alphas}a,
which consists of a single
struck quark and the proton remnant,
denoted by ``+1''.
At HERA energies there are significant higher-order
quantum chromodynamic (QCD) corrections:
to leading order in the strong coupling constant, $\alpha_{\rm s},$
these are QCD-Compton scattering (QCDC),
where a gluon is radiated by the
scattered quark and boson-gluon-fusion (BGF),
where the virtual boson and a gluon fuse to
form a quark-antiquark pair.
Both processes have 2+1 partons in the final
state, as shown in fig.~\ref{fig:feyn_alphas}.
There also exists calculations for the higher,
next-to-leading (NLO) processes.

HERA provides a unique opportunity to study the scale behaviour
of Quantum Chrmodynamics in a single experiment with a 
cleaner background environment than that obtainable at hadron colliders.
The data presented here were taken in 1994 and onwards
at the $ep$
 collider HERA using the H1 and ZEUS detectors. During this period HERA operated
with positrons of energy $E_e=27.5$~GeV and protons with energy
$820$~GeV.
A detailed descriptions of the H1 and ZEUS detectors can be found
in refs.~\cite{H1app} and~\cite{b:Detector} respectively.

\section{Structure functions}
\begin{figure}[hbt]
\centerline{\epsfig{file=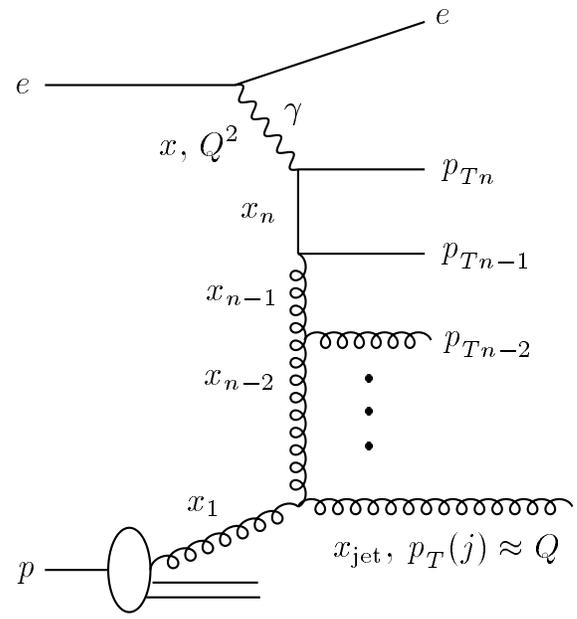,height=0.5\textwidth}}
\caption{Parton ladder diagram in DIS.}
\label{fig:ladder}
\end{figure}

Perturbative QCD (pQCD) does not predict the absolute value of the parton
densities within the proton but determines how they vary from a
given input. For a given initial distribution at a particular scale,
$Q_0^2,\ $
Altarelli-Parisi (DGLAP) evolution~\cite{dglap} enables the
distributions at higher $Q^2$ to be determined.
DGLAP evolution resums the leading
$\log(Q^2)$ contributions associated with a chain of gluon emissions.
At large enough positron-proton centre-of-mass energies there is a second
large variable $1/x$ and, therefore, it is also necessary
to resum the $\log(1/x)$ contributions.
This is achieved by using the BFKL equation~\cite{bfkl}.
In the DGLAP parton evolution scheme
the parton cascade follows a strong ordering in transverse
momentum
$p_{Tn}^2 \gg p_{Tn-1}^2 \gg... \gg p_{T1}^2$,
while there is only a soft
(kinematical) ordering for the fractional momentum
$x_n<x_{n-1}<...<x_1$ (see figure~\ref{fig:ladder}.)
By contrast, in the BFKL
 scheme the cascade follows a strong ordering in fractional
momentum
$x_n \ll x_{n-1} \ll... \ll x_1$,
while there is no ordering in transverse
momentum.

At small $x$ the dominant parton is the gluon and the description of 
the structure function is driven by the behaviour of the gluon.
Because of gluon splitting, $g \rightarrow q\bar q,$
pQCD suggests the small $x$ behaviour of the
sea quark and gluon distributions are strongly correlated.

\begin{figure}[htb]
\centerline{\epsfig{file=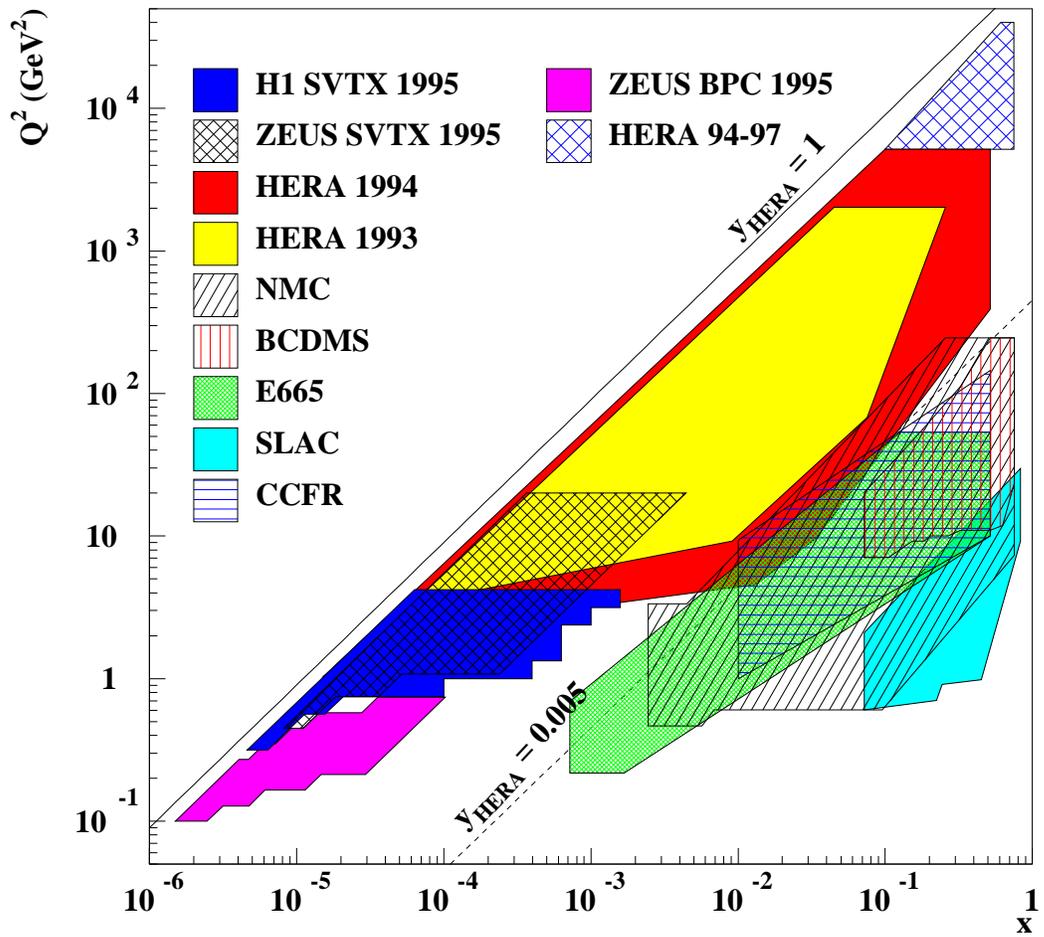,height=0.75\textwidth,bbllx=5pt,bblly=10pt,bburx=560pt,bbury=520pt}}
\caption{The kinematical region covered by HERA and fixed target
experiments}
\label{fig:kinem_plane}
\end{figure}

The kinematic plane covered by HERA and the fixed target measurements is
shown in fig.~\ref{fig:kinem_plane}. HERA has increased the reach in
$Q^2$ by about 2 orders of magnitude and can also probe nearly 3 orders
of magnitude further down in $x$. 
The low $x$ region is correlated with low values
of $Q^2.$ The differential NC DIS
cross section is related to three structure functions:

\begin{equation}
\frac{d^2\sigma^{e^{\pm}p}}{dxdQ^2}  = 
\frac{2\pi\alpha^2}{xQ^4}(Y_+F_2(x,Q^2)
   - y^2F_L(x,Q^2) \mp Y_-xF_3(x,Q^2)),
\end{equation}

\noindent where $Y_{\pm} = 1 \pm (1-y)^2.$ The structure function
$F_2$ in QPM is
just the sum of the quark densities multiplied by the appropriate
electric charge;
$F_3$ arises from the weak part of the cross section
and is negligible for $Q^2 < 5000 {\rm \ GeV^2,}$ and 
$F_L$ is the longitudinal structure function and only
becomes important for $ y > 0.6.$ Hence by measuring the differential
cross section at HERA one is effectively measuring the structure
function $F_2.$

\begin{figure}[p]
\centerline{\epsfig{file=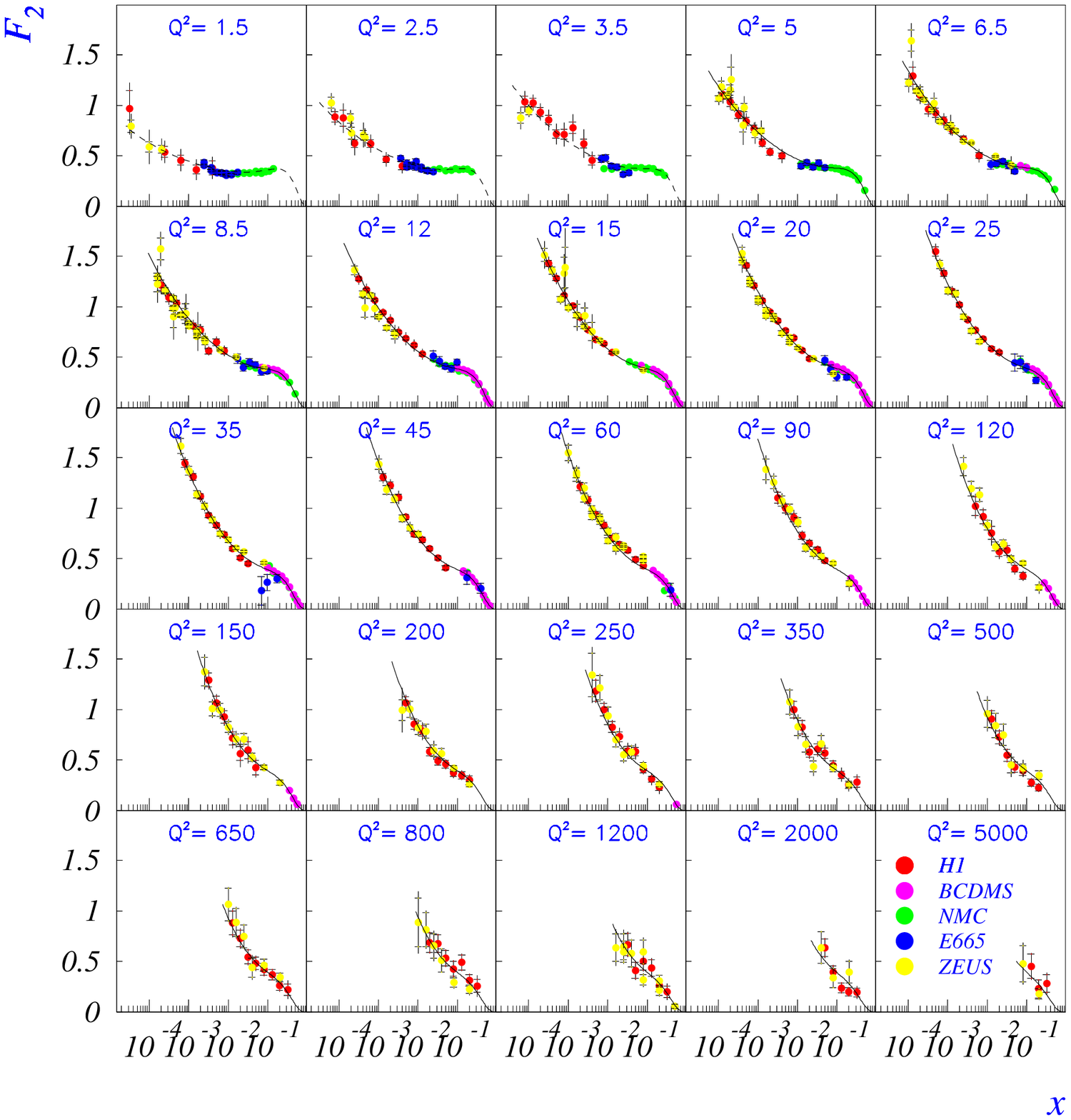,width=.95\textwidth,bbllx=25pt,bblly=140pt,bburx=560pt,bbury=695pt}}
\caption{Measurement of $F_2(x,Q^2)$ as function of $x.$ }
\label{fig:f2x}
\end{figure}

\begin{figure}[p]
\centerline{\epsfig{file=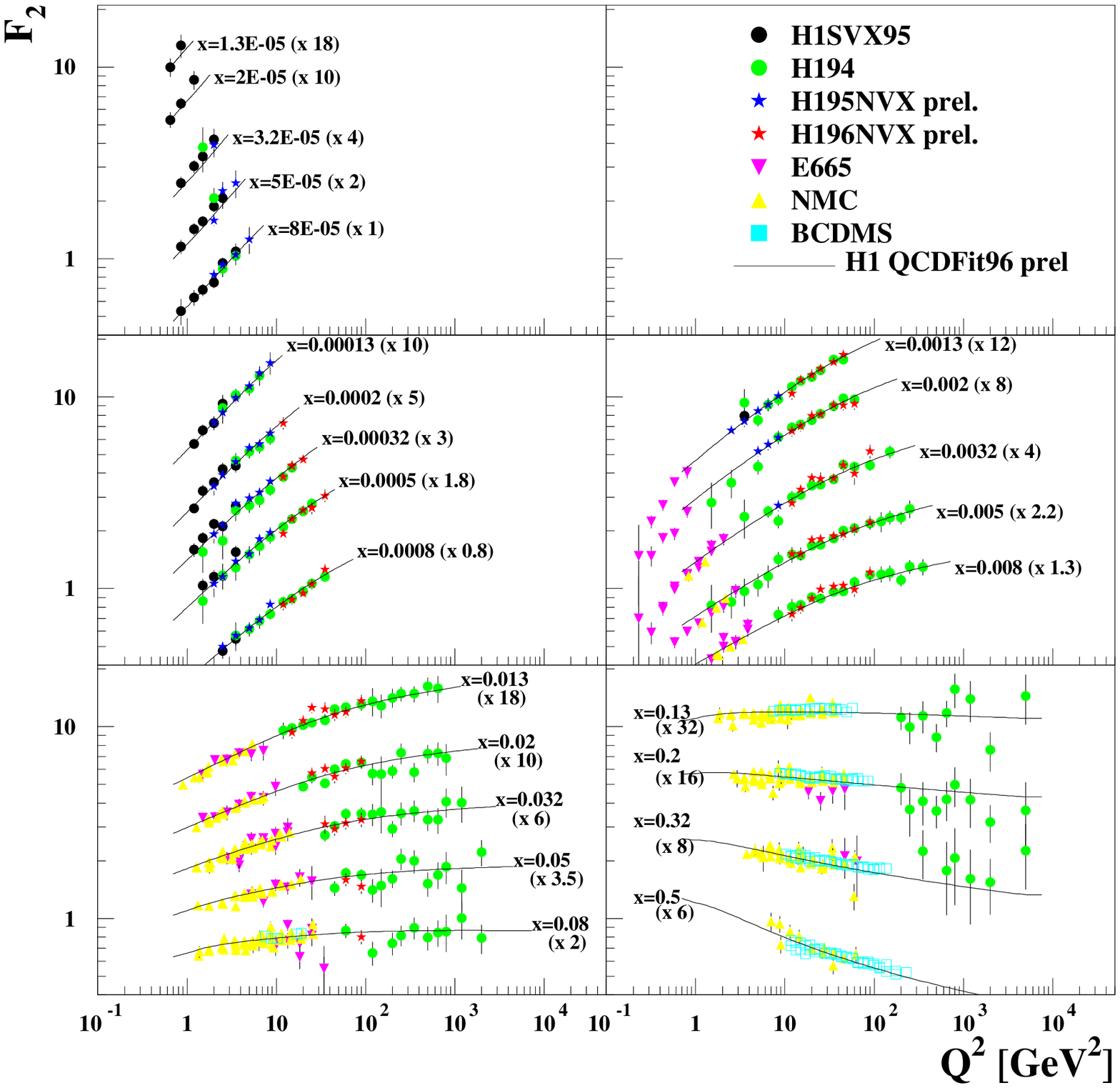,width=.95\textwidth,bbllx=10pt,bblly=5pt,bburx=600pt,bbury=575pt}}
\caption{Measurement of $F_2(x,Q^2)$ as function of $Q^2.$}
\label{fig:f2q2}
\end{figure}

The $F_2$ measurements are shown in figs.~\ref{fig:f2x} and
\ref{fig:f2q2} as
a function of $x$ and $Q^2$ respectively. The error bars are at the
5-10\% level and the normalisation uncertainty is $\sim$2\%.
There is a steep rise of $F_2$ with decreasing $x$ in all $Q^2$ bins,
fig.~\ref{fig:f2x}.
Scaling violations in $Q^2$
are clearly seen in fig.~\ref{fig:f2q2}.
Both H1 and ZEUS have performed next to leading order (NLO) 
QCD fits~\cite{h1fit,zfit}
based on the DGLAP evolution equations using both HERA and fixed target
data. Fig.~\ref{fig:f2q2} shows that these QCD fits describe the $F_2$
data well, though it should be noted that
the data can also be satisfactorily described by the BFKL
prediction~\cite{F2BFKL}.

\begin{figure}[htb]
\centerline{\epsfig{file=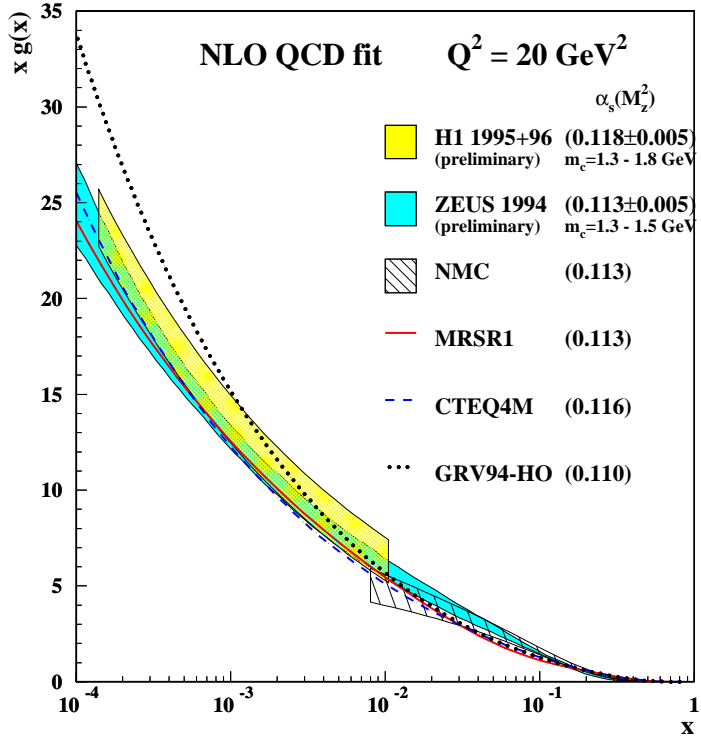,height=10.0cm,bbllx=50pt,bblly=190pt,bburx=500pt,bbury=665pt}}
\caption{The gluon density $xg(x)$ at $Q^2 = 20 {\rm \ GeV^2}$ extracted
from NLO QCD fits by the H1, ZEUS and NMC collaborations.}
\label{fig:gluon}
\end{figure}

The scaling violations from the HERA data allow an estimate of the gluon
density $xg(x)$ at low values of $x,$ whilst the fixed target data are
used to constrain the high $x$ region.
The extracted gluon densities from the fits are shown in
fig~\ref{fig:gluon} for a fixed $Q^2 = 20 {\rm \ GeV^2.}$ The error band
shows the statistical and systematic uncertainty taking into account 
correlations and variations in the mass of the charm quark, $m_c,$
and the strong coupling constant, \als. The results of the two HERA experiments
are in good agreement and the extracted densities agree with the results of
NMC~\cite{NMC} for large $x$. 
The resulting gluon distributions show a clear rise with
decreasing $x$ and have a
15\% uncertainty at $x \sim 5\times 10^{-4}.$
These NLO QCD fits are also in good agreement with the global QCD
analyses performed by MRS~\cite{MRS} and CTEQ~\cite{CTEQ}, 
whilst the prediction from the
dynamical evolution of GRV~\cite{GRV} is too steep for $x < 10^{-3}.$

\section{Forward Jet Production in DIS}
The $F_2$ measurements fail to distinguish between DGLAP and the BFKL
approach to the QCD evolution. The hadronic final states are expected to
give additional information. For events at low $x,$ hadron production
between the current jet and the proton remnant is expected to be
sensitive to the effects of BFKL or DGLAP dynamics.
A possible signature of BFKL dynamics is the behaviour of DIS events
at low $x$
which contain a jet that has a transverse momentum $p_T^2(j) \approx Q^2$
(so minimizing the phase space available for DGLAP evolution) and
has longitudinal momentum fraction (of the proton) $x_{jet}$ that
is large (in order to maximise the phase space for BFKL evolution),
fig~\ref{fig:ladder}.

\begin{figure}[hbt]
\centering
\mbox{
\subfigure[]
{\epsfig{file=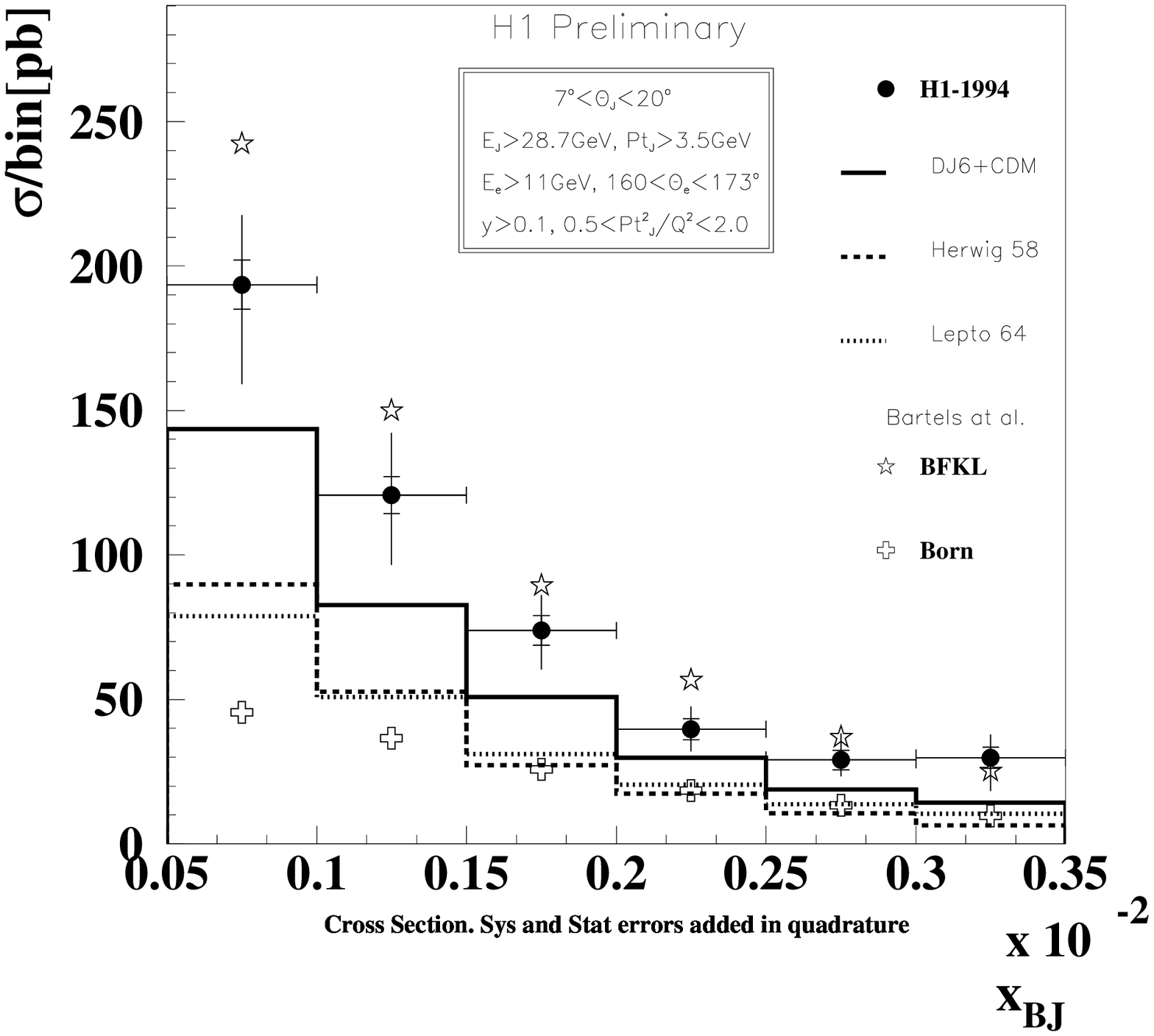,width=0.5\textwidth,bbllx=5pt,bblly=0pt,bburx=560pt,bbury=500pt}}
\subfigure[]
{\epsfig{file=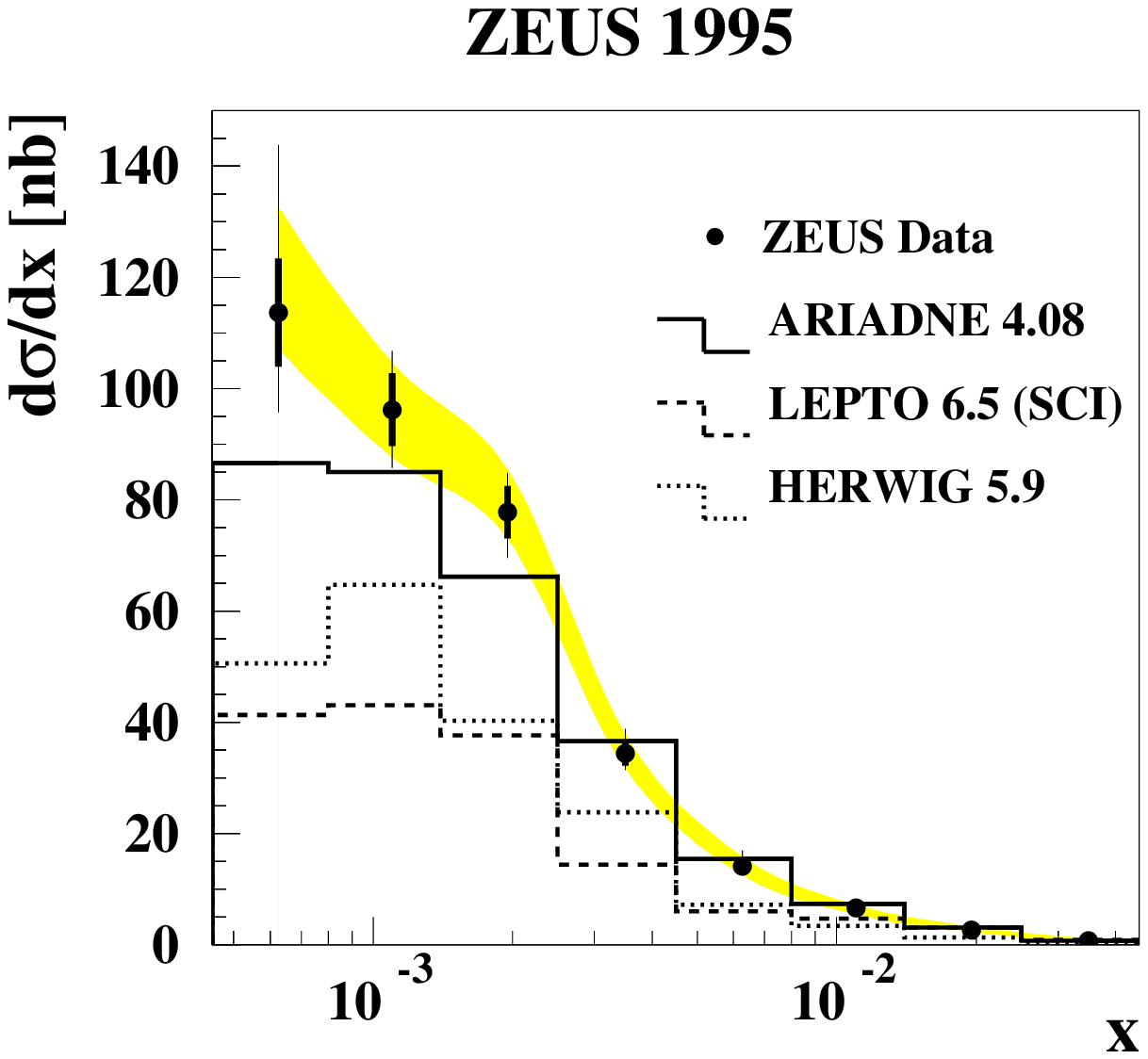,width=.5\textwidth,bbllx=10pt,bblly=75pt,bburx=375pt,bbury=405pt}}}
\caption{Measurement of forward jet production as function of $x$ by (a)
H1 and (b) ZEUS experiments. The errors due to the uncertainty in the
jet energy scale are show in the shaded band.}
\label{fig:fjets}
\end{figure}

In Fig.~\ref{fig:fjets}, recent data from H1~\cite{H1-forward} and
ZEUS~\cite{riveline} are compared with BFKL predictions~\cite{bartelsH1}
and fixed order QCD predictions as calculated with the
MEPJET~\cite{mepjet}
program at NLO. The conditions $p_T(j)\simeq Q$ and
$x_{jet}\gg x$ are satisfied in the two experiments by slightly
different
selection cuts. H1 selects events with a forward jet of $p_T(j)>3.5$~GeV
(in the angular region $7^o < \theta(j) < 20^o$) with
\begin{equation}
   0.5  <  p_T(j)^2/Q^2\; < \; 2\;, \qquad \qquad
   x_{jet}  \simeq  E_{jet}/E_{proton} > 0.035\;; \label{eq:fj-H1}
\end{equation}
while ZEUS triggers on somewhat harder jets of $E_T(j)>5$~GeV
and $\eta(j)<2.6$ with
\begin{equation}
   0.5  <  p_T(j)^2/Q^2\; < \; 2\;, \qquad \qquad
   x_{jet}  =  p_z(j)/E_{proton} > 0.036\;. \label{eq:fj-ZEUS}
\end{equation}

Fig.~\ref{fig:fjets} shows that
both experiments observe a forward jet cross section which rises steeply
with decreasing $x$ with substantially more forward jet events
than expected from NLO QCD (labelled as Born in fig.~\ref{fig:fjets}a.) 
A BFKL calculation (the stars) gives a better agreement
with the data.
The overall normalisation in this calculation
is uncertain and the agreement may be
fortuitous. Indeed, it should also be noted
that both experiments observe more centrally
produced dijet events than predicted by the NLO QCD calculations.
The ARIADNE Monte Carlo (CDM) model
describes the
steeply increasing jet cross section with decreasing $x$. 
The ARIADNE model does not have a strong ordering in transverse
momentum in the QCD cascade, akin to BFKL type dynamics, although it does
not make explicit use of the BFKL equation.
Whilst those
models that adhere to the DGLAP formalism (LEPTO and HERWIG) fail to
predict this large growth.
Further careful investigation is necessary before claiming that BFKL is
the mechanism for this enhanced forward jet production.

\section{Fragmentation Functions}
Fragmentation functions represent the
probability
for a parton to fragment into a particular
hadron carrying a certain fraction of the parton's energy
and, like structure functions,
cannot be calculated in perturbative QCD, but
can be evolved from a starting distribution at a defined energy scale.
If the fragmentation functions are combined with
the cross sections for the inclusive production of
each parton type in the given physical process, predictions can be made
for the scaled momentum, $x_p,$ spectra of final state hadrons.
Small $x_p$ fragmentation is significantly affected by the coherence
(destructive interference) of
soft gluons~\cite{basics}, whilst
scaling violation of the fragmentation function at
large $x_p$ allows a measurement of \als~\cite{webnas}.

A natural frame in which to study the dynamics of the hadronic final
state
in DIS is the Breit frame~\cite{feyn}.
In this frame the exchanged
virtual boson is purely
space-like with 3-momentum ${\bf q}=(0,0,-Q)$, the incident quark
carries momentum $Q/2$ in the positive $Z$ direction,
and the outgoing struck quark carries Q/2 in the negative
$Z$ direction.  A final state particle has a 4-momentum
$ p^B$ in this frame,
and is assigned to the current region if $p^B_Z$ is negative, and to the
target frame if $p^B_Z$ is positive.
The advantage of this
frame lies in the maximal separation of the outgoing parton from
radiation associated with the incoming parton and the proton remnant,
thus providing the optimal environment for the study of the
fragmentation of the outgoing parton.

In \ee~annihilation the two quarks are produced
with equal and opposite momenta, $\pm \sqrt{s}/2.$
This can be compared with
a quark struck from within the
proton with outgoing momentum $-Q/2$ in the Breit frame.
In the direction of the struck quark (the current fragmentation region)
the particle momentum spectra, $x_p = 2p^B/Q,$
are expected to have a
dependence on $Q$ similar to those observed in
\ee~annihilation~\cite{eedis,anis,char} at energy $\sqrt{s}=Q.$

\begin{figure}[hb]
\centerline{\epsfig{file=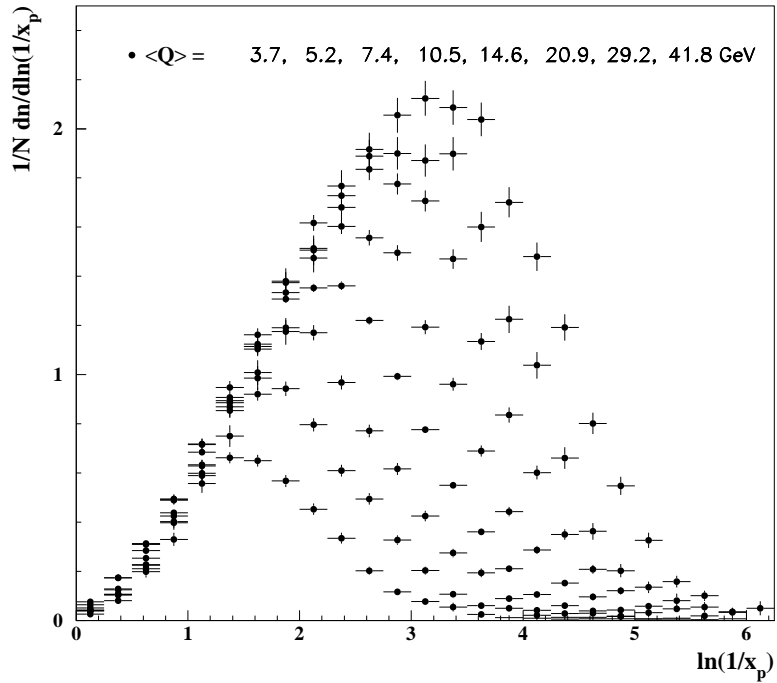,width=0.65\textwidth}}
\caption{Preliminary ZEUS results of 
the evolution of the $1/N dn/d\log(1/x_p)$ distributions with $Q^2.$
Only statistical errors are shown.}
\label{fig:logxp}
\end{figure}

\begin{figure}[bt]
\centerline{\epsfig{file=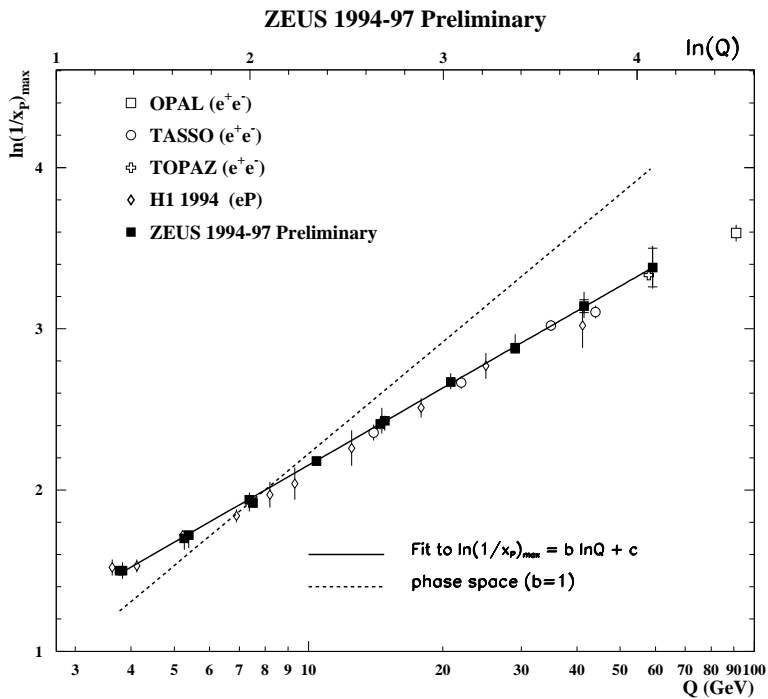,width=0.65\textwidth}}
\caption{Preliminary ZEUS results of 
evolution of the peak position \logxpmax~with $Q^2.$}
\label{fig:peak}
\end{figure}

In fig~\ref{fig:logxp} the
\logxp~distributions for charged particles in the current fragmentation
region of the Breit frame are shown as a function
of $Q^2.$
These distributions are approximately Gaussian in shape with
mean charged multiplicity given by the integral of the distributions.
As $Q^2$ increases the multiplicity increases and the the peak of the
distributions shifts to larger values of \logxp.
Figure~\ref{fig:peak} shows
this peak position, \logxpmax, as a function of $Q$
for the HERA data and of $\sqrt{s}$ for the $e^+e^-$ data.
Over the range shown
the peak moves from $\simeq$~1.5 to 3.3.
The HERA data points are
consistent with those from TASSO and TOPAZ
and a clear agreement in the rate of
growth of the HERA points with
the $e^+e^-$ data at higher $Q$ is observed.

The increase of \logxpmax~can be approximated phenomenologically
by the straight line fit
$ \logxpmax = b \log(Q)+c $
also shown in figure~\ref{fig:peak}.
The values obtained from the fit to the ZEUS data are
$b=0.69 \pm 0.01{\rm (stat)} \pm 0.03{\rm (sys)}$ and
$c=0.56 \pm 0.02 ^{+0.08}_{-0.09}.$
The gradient
extracted from the OPAL and TASSO data is $b=0.653\pm0.012$ (with
$c=0.653\pm0.047$) which is consistent with the ZEUS result. This value
is consistent with that published
by OPAL, $b=0.637\pm 0.016$,
where the peak position was extracted using an
alternative method~\cite{opal}.
A consistent value of the gradient is therefore determined in DIS and
\ee~annihilation experiments.

Also shown is the statistical fit to the data
when $b=1$ ($c=0.054 \pm 0.012$) which would be the case if the QCD
cascade
was of an incoherent nature, dominated by cylindrical phase space.
The observed gradient is clearly inconsistent with $b=1$
and therefore inconsistent with cylindrical phase space.

\begin{figure}[tb]
\centerline{\epsfig{file=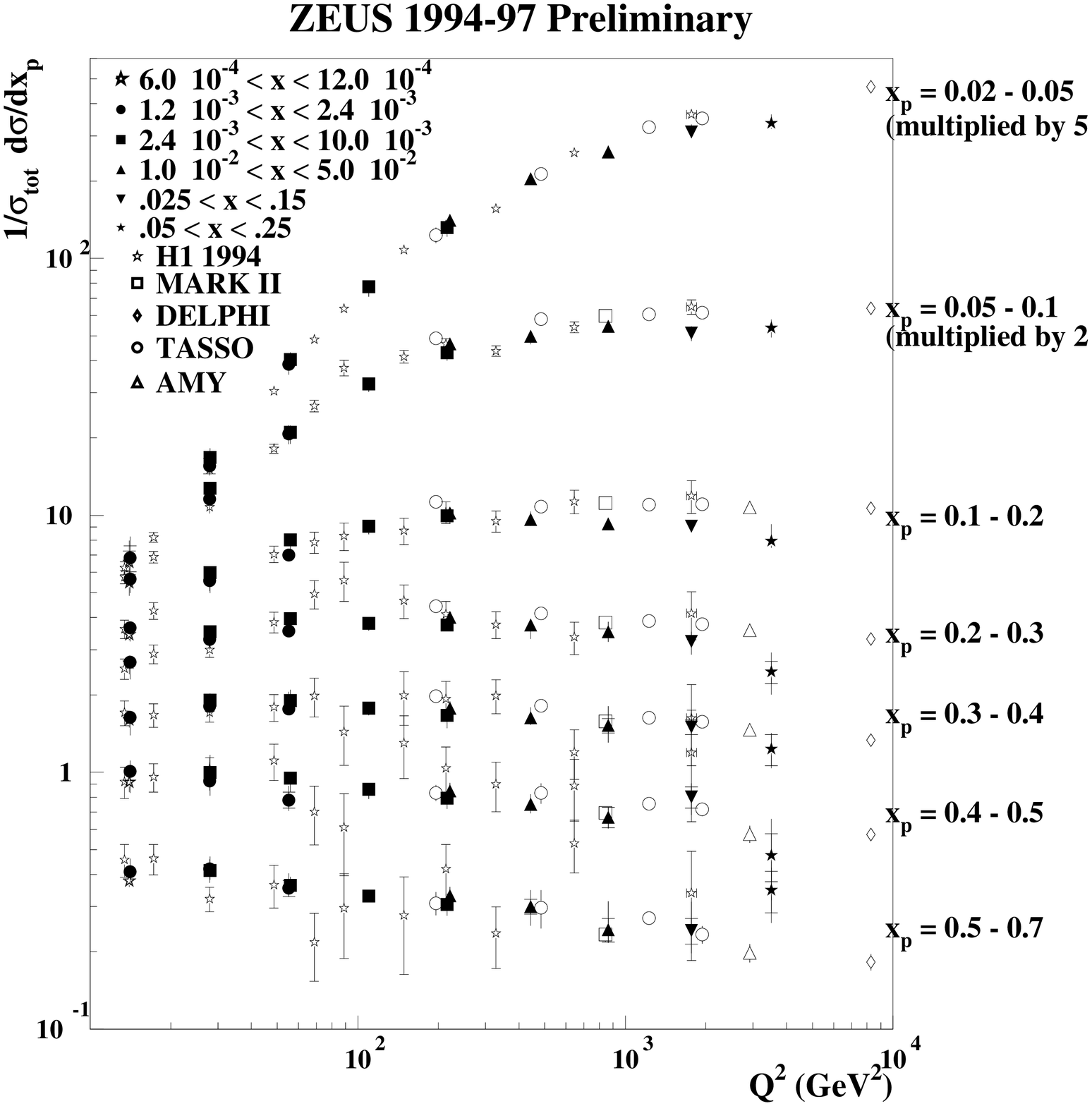,width=0.95\textwidth}}
\caption{\small The inclusive charged particle distribution,
$ 1/\sigma_{tot}~ d\sigma/dx_p$,
in the current fragmentation region of the Breit frame as measured by H1
and ZEUS collaborations.
The open points represent data from $e^+e^-$ experiments divided by two
to
account for $q$ and $\bar q$ production (also corrected
for contributions from $K^0_S$ and $\Lambda.$)}
\label{fig:largexp}
\end{figure}

The inclusive charged particle distribution,
$ 1/\sigma_{tot}~ d\sigma/dx_p$,
in the current fragmentation region of the Breit frame are shown in bins
of $x_p$ and $Q^2$ in fig.~\ref{fig:largexp}.
The increasingly steep fall-off, at fixed $Q^2,\ $ towards
higher values of $x_p$ as $Q^2$ increases, shown in
figure~\ref{fig:largexp},
corresponds to  the production of
more particles with
a smaller fractional momentum, and
 is indicative of scaling
violation in the fragmentation function.
For $Q^2 > 80{\rm\ GeV^2}$ the distributions rise
with $Q^2$ at low $x_p$ and
fall-off at high $x_p$ and high $Q^2$.
In  figure~\ref{fig:largexp} the HERA data are compared at $Q^2=s$
to $e^+e^-$ data~\cite{eedata},
again divided by two to account for the production of both a $q$ and $\bar q.$
In the $Q^2$ range shown there is good agreement between the
current region of the Breit frame in DIS and the $e^+e^-$
experiments.

\section{Summary}
This review gives a brief summary of a small sample
of the QCD results coming from HERA. 
Charged particle distributions have been studied in the current region
of the Breit frame over a wide range of $Q^2.$ These result show clear
evidence for scaling violation in scaled momenta as a function of $Q^2$
and supports the coherent nature of QCD cascades. The observed charged
particle spectra are consistent with the universality of quark
fragmentation in
$e^+e^-$ and DIS.

The intriguing rise of $F_2$
at small $x$ can be well described using conventional DGLAP evolution
equations. The data though can also be described by the BFKL approach
thus giving rise to ambiguities how to treat QCD in this small $x$
regime. In order to resolve these ambiguities jet production in the
forward direction has been studied and the cross
section for such jets is seen not to be reproduced
by Monte Carlo models based on DGLAP parton shower evolution.

\newpage 

\newpage

\newpage

\newpage

\begin{thebibliography}{25}
\bibitem{DA} S.~Bentvelsen, J.~Engelen and P.~Kooijman,
Proceedings of the 1991 Workshop on Physics at HERA, DESY
Vol.~1~(1992)~23.

\bibitem{H1app}
H1 Collaboration, I.\ Abt et al., DESY preprint 93-103,
{\it Nucl.\ Instr.\ Meth.\ }{\bf 386}, 310  (1997) (Vol 1) and {\it ibid.} p.348 (Vol 2).

\bibitem{b:Detector} ZEUS Collab., The ZEUS Detector,
                     Status Report 1993, DESY 1993.

\bibitem{dglap}
G. Altarelli and G. Parisi, {\it Nucl.\ Phys.\ } {\bf126}, 297 (1977);
V.N. Gribov and  L.N. Lipatov,
{\it Sov.\ J.\ Nucl.\ Phys.\ }{\bf 15}, 438 and 675 (1972);
Yu. L. Dokshitzer, {\it Sov.\ Phys.\ JETP}{\bf 46}, 641 (1977).

\bibitem{bfkl}
E.A. Kuraev, L.N. Lipatov and V.S. Fadin,
{\it Sov.\ Phys.\ JETP} {\bf 45}, 199 (1977);
Y.Y. Balitsky and L.N. Lipatov,
{\it Sov.\ J.\ Nucl.\ Phys.\ }{\bf 28}, 282 (1978).

\bibitem{h1fit}
H1 collab., Paper 262 presented at Int.\ Europhysics Conf.\ on High
Energy Physics, HEP97, Jerusalem.

\bibitem{zfit}
M.\ A.\ J.\ Botje, Proc.\ of 5th Int.\ workshop on DIS and QCD (1997), 
(eds. Repond and Krakauer).

\bibitem{F2BFKL}
A.\ J.\ Askew {\it et al.\,} {\it Phys.\ Rev.\ }{\bf D47}, 3775 (1993);
{\bf D49}, 4402 (1994).

\bibitem{NMC}
NMC collab., M.\ Arneodo {\it et al.\,} {\it Phys.\ Lett.\ }{\bf B309},
222 (1993).

\bibitem{MRS}
A.\ Martin, R.\ Roberts and W.\ J.\ Stirling, {\it Phys.\ Lett.\ }{\bf
B387}, 419 (1996).

\bibitem{CTEQ}
H.\ L.\ Lai {\it et al.}, {\it Phys.\ Rev.\ }{\bf D55}, 1280 (1997).

\bibitem{GRV}
M.\ Gl\"uck, E.\ Reya and A.\ Vogt,{\it Z.\ Phys.\ }{\bf C67}, 433 (1995).

\bibitem{H1-forward}
H1 collab., paper pa03-049 submitted to 28th Int.\ Conf.\ on High Energy
Physics ICHEP'96, Warsaw, Poland.

\bibitem{riveline}
ZEUS collab., J.Breitweg {\it et al.\,} DESY preprint 98-050.

\bibitem{mepjet} 
E. Mirkes and D. Zeppenfeld,
{\it Phys.\ Lett.\ }{\bf B380}, 205 (1996).

\bibitem{bartelsH1}
J.\ Bartels {\it et al.\,} {\it Phys.\ Lett.\ } {\bf B384}, 300 (1996).
\bibitem{basics} Yu.~Dokshitzer, V. Khoze, A. Mueller and S. Troyan,
``Basics of Perturbative QCD'', Editions Fronti\`{e}res, Gif-sur-Yvette,
France (1991).

\bibitem{webnas} G.\ Altarelli {\it et al.}, {\it Nucl.\ Phys.\ }{\bf B160},
301 (1979);
P.~Nason and B.~R.~Webber, {\it Nucl.\ Phys.\ }{\bf B421}, 473 (1994).

\bibitem{feyn} R.P. Feynman, ``Photon-Hadron Interactions'', Benjamin,
N.Y.  (1972).

\bibitem{eedis} Yu.~Dokshitzer {\it et al.}, {\it Rev.\ Mod.\ Phys.\ }{\bf 60},
373 (1988).

\bibitem{anis} A. V. Anisovich {\it et al.}, {\it Il Nuovo Cimento }{\bf A106},
547 (1993).

\bibitem{char} K.\ Charchu{\l}a, {\it J.\ Phys.\ } {\bf G19}, 1587 (1993).

\bibitem{opal} OPAL Collab., M.~Akrway {\it et al.}, {\it Phys. Lett.\ }
{\bf B247}, 617 (1990). 

\bibitem{topaz}
TOPAZ collab., R.\ Itoh {\it et al.},{\it Phys.\ Lett.\ }{\bf B345}, 335 (1995).

\bibitem{eedata} 
TASSO Collab., W.~Braunschweig {\it et al.},{\it Z.\ Phys.\ }{\bf C47}, 187 (1990);
MARK II Collab., A.~Petersen {\it et al.}, {\it Phys.\ Rev.\ }{\bf D37}, 1 (1988);
AMY Collab., Y.~K.~Li {\it et al.}, {\it Phys.\ Rev.\ }{\bf D41}, 2675 (1990);
DELPHI Collab., P.~Abreu et {\it al.}, {\it Phys.\ Lett.\ }{\bf B311}, 408 (1993).
\end{thebibliography}
\end{document}